\documentclass[sn-mathphys-num, iicol]{sn-jnl}

\usepackage{graphicx}%
\usepackage{multirow}%
\usepackage{amsmath,amssymb,amsfonts}%
\usepackage{amsthm}%
\usepackage{mathrsfs}%
\usepackage[title]{appendix}%
\usepackage{xcolor}%
\usepackage{textcomp}%
\usepackage{manyfoot}%
\usepackage{booktabs}%
\usepackage{algorithm}%
\usepackage{algorithmicx}%
\usepackage{algpseudocode}%
\usepackage{listings}%

\usepackage{lineno}

\usepackage[colorinlistoftodos]{todonotes}

\begin{document}


\title{Scaling laws for softened hadron production at LHC energies}


\author[1]{\fnm{D.} \sur{Rosales Herrera}}

\author[1]{\fnm{J. R.} \sur{Alvarado García}}

\author[1]{\fnm{A.} \sur{Fernández Téllez}}

\author[1]{\fnm{E.} \sur{Cuautle}}

\author*[3]{\fnm{J. E.} \sur{Ramírez}}\email{jhony.eredi.ramirez.cancino@cern.ch}

\affil[1]{Facultad de Ciencias F\'isico Matem\'aticas, Benem\'erita Universidad Aut\'onoma de Puebla,
Apartado Postal 165, 72000 Puebla, Puebla, Mexico}

\affil[2]{Instituto de Ciencias Nucleares, Universidad Nacional Autónoma de México,
Apartado Postal 70-543, Ciudad de México 04510, México}

\affil[3]{Centro de Agroecología, Instituto de Ciencias, Benemérita Universidad Autónoma de Puebla, Apartado Postal 165, 72000 Puebla, Puebla, M\'exico}


\abstract{In this paper, we conduct a data-driven study of the production of softened hadrons and their contribution to the transverse momentum spectrum. 
To this end, we assume that the production of charged particles at soft and hard scales fundamentally results from the fragmentation of color strings.
We analyze the \textit{p}$_\text{T}$-spectrum data from pp to AA collisions at LHC energies reported by the ALICE Collaboration, finding that, in all cases, the data can be collapsed into a \textit{p}$_\text{T}$-exponential trend in the range 1 GeV$<$\textit{p}$_\text{T}<$6 GeV.
With this insight, the description of the \textit{p}$_\text{T}$-spectrum should contain information on the charged particle production coming from two different sources: fragmentation of color strings and collective phenomena that redistribute the transverse momentum and enhance the production of particles at intermediate \textit{p}$_\text{T}$.
We also found different relations between the effective temperature, multiplicity, and average \textit{p}$_\text{T}$ for pp and AA collisions, indicating inherent dissimilarities between small and large colliding systems. In contrast, the contribution of the softened hadrons to the \textit{p}$_\text{T}$-spectrum and average \textit{p}$_\text{T}$ collapse onto scaling laws. Our results show that the physical mechanisms producing softened hadrons have similar origins for all colliding systems, revealing a stronger dependence on freeze-out parameters rather than the system size. 
}

\maketitle

\section{Introduction}\label{sec:intro}

Experimental evidence of the quark-gluon plasma (QGP) formation in heavy ion collision shed light on the properties of the strongly interacting matter under extreme conditions of temperature and pressure \cite{BRAHMS:2004adc,PHOBOS:2004zne,STAR:2005gfr,PHENIX:2004vcz}.
The created system efficiently transforms initial spatial anisotropies into correlated momentum disparities among the resultant particles, generating the well-known collective flow \cite{POLLERI199819}.
In particular, AA collisions show a radial flow affecting the $p_\text{T}$-spectrum of produced particles, which is more relevant for heavy hadrons at intermediate $p_\text{T}$ as the behavior of the proton-to-pion ratio reveals \cite{ALICE:2019hno}. Other mechanisms similarly modifying the $p_\text{T}$-spectrum are the suppression of high $p_\text{T}$ hadrons, parton recombination, color reconnection, or multiple parton interactions \cite{Das:1977cp,Hwa:2003ic,Cuautle:2015fbx}.
The hydrodynamical model can also explain the flow's origin by assuming a medium's presence \cite{Belenkij:1955pgn}. The energy loss due to the parton scattering and bremsstrahlung can suppress the production of high $p_\text{T}$ hadrons \cite{Baier:2000mf,Zakharov:2007pj}, leading to the observed jet quenching \cite{CMS:2011iwn}.
These results highlight that a fraction of the produced hadrons at the initial state get softened through the system's evolution \cite{PHENIX:2003djd}. 

Similar effects to collective phenomena and radial flow have been observed 
in events with the largest multiplicities in pp and pPb collisions \cite{ALICE:2013wgn,CMS:2016fnw,ALICE:2018pal,ALICE:2020nkc}, where the softening of high $p_\text{T}$ hadrons is less noticeable than in the heavy ion case.
Recent discussions about the formation of QGP droplets motivate the search for jet quenching-like effects in small systems \cite{PHENIX:2018lia,Nagle:2018nvi}. The study of the flow patterns and the strong differences between small and large systems motivates the development of models describing particle production in ultrarelativistic collisions \cite{Cuautle:2007im}.
For instance, the fragmentation of color strings emerging from the colliding partons describes the particle production at the parton level \cite{andersson1998lund}. 
This phenomenological approach is part of event generators, such as PYTHIA, EPOS, and Herwig, which consider the string fragmentation to describe experimental observables as well as the production of particles at high energies \cite{Andersson:1983ia}. 
A fundamental aspect of this framework is incorporating the Schwinger mechanism to describe the $p_\text{T}$-distribution of the charged particles created. 
The Schwinger mechanism can adequately describe the $p_\text{T}$-spectrum data by considering a stochastic picture of the string tension \cite{BIALAS1999301,Herrera:2024zjy}. 
In particular, it is possible to reproduce the asymptotic behaviors experimentally observed by assuming a heavy-tailed description for the string tension fluctuations \cite{Pajares:2022uts,Garcia:2023eqg,Herrera:2024zjy,Herrera:2024tyq}.  
An immediate consequence of this approach is that the systems formed in ultrarelativistic collisions are no longer thermal, emerging nonequilibrium temperature fluctuations along the systems \cite{Herrera:2024tyq}. 
Nevertheless, for high multiplicity events, the aforementioned collective phenomena redistribute the $p_\text{T}$ of a portion of the produced hadrons, enhancing the $p_\text{T}$-spectrum at intermediate values (1–6 GeV). 
In this way, the fragmentation of color strings fails to capture the softened hadron production.

This paper aims to describe the entire $p_\text{T}$-spectrum of charged particles produced in ultrarelativistic collisions. In particular, we are interested in identifying the softened hadron production and quantifying their contribution to the freeze-out parameters. 
To this end, we analyze the experimental $p_\text{T}$-spectrum data at intermediate $p_\text{T}$ values by assuming a color string fragmentation baseline for charged particle production. 
This procedure allows us to data-driven infer the functional form of the softened hadron contribution to the $p_\text{T}$-spectrum, which is now compounded by the sum of the particle production coming from string fragmentation and softened hadrons. Then, we can determine the contribution of the softened hadrons to the freeze-out parameter, finding scaling laws in terms of the multiplicity rather than the system’s size.

The rest of the manuscript is organized as follows. In Sec.~\ref{sec:TMD}, we use a nonextensive string fragmentation framework to describe the $p_\text{T}$ spectrum of the charged particles produced in high-energy collisions. This approach correctly reproduces the asymptotic behaviors experimentally observed. We use this approach as a baseline for the particle production at the soft and hard scales. 
In Sec.~\ref{sec:elucidating}, we data-driven infer that softened hadrons mainly contribute to the $p_\text{T}$-spectrum at the intermediate region, following an exponential decay independently of the system's size, center of mass energies, or event classifications. Consequently, the softened particles are produced at a higher temperature than the soft scale.
With this insight, in Sec.~\ref{sec4:complete}, we describe the entire $p_\text{T}$ spectrum by considering that the charged particle production comes from combining the color string fragmentation and collective phenomena.
The latter redistributes the $p_\text{T}$ of a fraction of particles, leading to the softened hadrons.
In Sec.~\ref{sec:contribution}, we quantify the contribution of the softened hadrons to the $p_\text{T}$-spectrum, the multiplicity, and $\langle p_\text{T} \rangle$, finding scaling laws as function of the the freeze-out parameters.  Finally, we present our final remarks and conclusions in Sec.~\ref{sec:conclu}.

\section{$p_\text{T}$-spectrum from a nonextensive description}
\label{sec:TMD}
String models describe particle production by the fragmentation of color strings emerging from the colliding partons. The particle-antiparticle creation process involves the Schwinger mechanism, given by \cite{schwinger}
\begin{equation}
   \frac{dN}{dp_\text{T}^2} \sim e^{-\pi p_\text{T}^2/x^2}, 
\end{equation}
which is the $p_\text{T}$-distribution when the charged particles are created from the fragmentation of color strings with tension $x^2$. 
In a general scenario the string tension may fluctuate accordingly to a probability density function $P(x)$ \cite{BIALAS1999301,Herrera:2024zjy}. Then, the $p_{\text{T}}$-spectrum must be computed as the convolution 
\begin{equation}
    \frac{dN}{dp_\text{T}^2}\sim \int_0^\infty e^{-\pi p_\text{T}^2/x^2} P(x)dx.
\label{eq:convolution}
\end{equation}
For instance, the well-known thermal distribution 
$$\frac{dN}{dp_\text{T}^2}  \sim e^{-p_\text{T}/T_\text{th}},$$
with soft scale $T_\text{th}=\langle p_\text{T}^2 \rangle/2$, is obtained by considering Gaussian string tension fluctuations \cite{BIALAS1999301,Herrera:2024zjy}.
This approach can describe the experimental data at very low energy collisions or low $p_\text{T}$ values.
However, it fails to describe the case of high energy collisions at high $p_\text{T}$ values. 
In part, it is because of the probability of hard gluon emission from strings growing with the center of mass energy, as stated in the Lund model \cite{andersson1998lund}, resulting in additional string tension fluctuations that the Gaussian distribution can no longer describe.

To incorporate the hard part of the $p_\text{T}$-spectrum, we must raise the probability of observing strings with higher tension. This can be achieved by enhancing the tail of the Gaussian distribution by promoting it to a $q$-Gaussian distribution 
\begin{equation}
 P(x)=\mathcal{N}_q \left[1+ (q-1)x^2/2\sigma^2\right]^\frac{1}{1-q},
 \label{eq:qGauss}
\end{equation}
which is a heavy-tailed probability density function broadly used to study nonequilibrium systems \cite{budini2}. In Eq.~\eqref{eq:qGauss},
$\mathcal{N}_q$ is the normalization constant, $q$ is the deformation parameter quantifying the deviations from the Gaussian behavior, and $\sigma$ is related to the width of the distribution \cite{budini2}. In this context, $q$ must take a value between 1 and 3 to guarantee the convergence of the normalization constant and allow variations in the string tension across the range from zero to infinity \cite{budini,budini2,Garcia:2023eqg}.

\begin{figure*}
\centering
\includegraphics{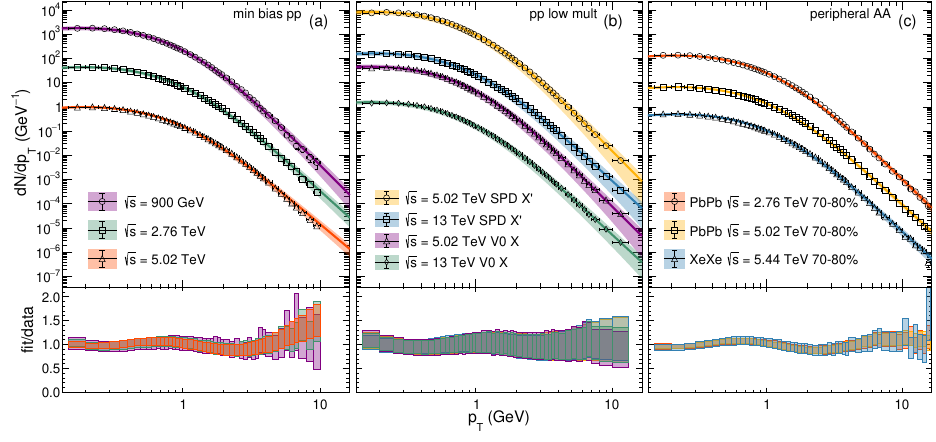}
\caption{Experimental
data of minimum bias (a), and low multiplicity events (b) in pp collisions, and peripheral events (c) in AA collisions at different center of mass energies reported
by the ALICE Collaboration accurately described by the Tricomi's function \eqref{eq:U}. 
    }
    \label{fig:U}
\end{figure*}

In this manner, by assuming a $q$-Gaussian description of the string tension fluctuations, the $p_\text{T}$-spectrum becomes a Tricomi's function 
\begin{equation}
 \frac{dN}{dp_\text{T}^2} \sim 
 U\left(a, 1/2, z_0 p_\text{T}^2\right),
 \label{eq:U}
\end{equation}
with $a=1/(q-1)-1/2$ and $z_0=\pi (q-1)/2\sigma^2$. 
Interestingly, this $U$ function has the experimentally observed asymptotic behaviors. At low $p_\text{T}$ values, $U$ behaves as the thermal distribution, i.e., $U \sim e^{-p_\text{T}/T_U}$ with soft scale $ T_U=  B(a ,1/2)/\sqrt{4\pi z_0}$, and $B$ being the beta function \cite{Pajares:2022uts}.
Conversely, it behaves as the power law $U \propto (p_\text{T}^2)^{\frac{1}{2}-\frac{1}{q-1}}$ for high $p_\text{T}$ values. In particular, this function adequately fits the $p_\text{T}$-spectrum data of the charged particle and Higgs boson production in pp collisions \cite{Pajares:2022uts,Garcia:2023eqg,Herrera:2024zjy, Herrera:2024tyq}. 
This is corroborated by analyzing the experimental data of minimum bias and low multiplicity events in pp collisions and peripheral events in AA collisions reported by ALICE Collaboration (see top panels of Fig.~\ref{fig:U}) \cite{ALICE:2010syw,ALICE:2018vuu,ALICE:2018hza,ALICE:2019dfi,ALICE:2022xip}.
One advantage of this approach is that the $U$ function controls the uncertainty propagation in all cases. Additionally, the fit to data ratio remains close to 1 for all the analyzed cases (see bottom panels of Fig.~\ref{fig:U}).

However, the $U$ function deviates from the $p_\text{T}$-spectrum data for the most central heavy ion collisions at intermediate $p_\text{T}$ values. In fact, these effects are more noticeable for high multiplicity events \cite{Garcia:2023eqg}.
It may occur because of an excess of charged hadrons detected with $p_\text{T}$ approximately from 1 to 6 GeV, usually explained by mechanisms beyond the string fragmentation, such as experimentally measured signals in AA collisions as well as similar signals modeled and observed in high multiplicity events in pp collisions \cite{Cuautle:2015fbx,Nagle:2018nvi}.
In this sense, we distinguish two contributions to the $p_\text{T}$-spectrum to describe the experimental data. 
The first assumes the production of charged particles through the fragmentation of color strings, described by the Tricomi's function \cite{Pajares:2022uts,Garcia:2023eqg,Herrera:2024zjy}. The second part considers
softened hadrons acquiring their transverse momentum through processes beyond string fragmentation, enhancing the $p_\text{T}$-spectrum at intermediate $p_\text{T}$ values. 

We must emphasize that $dN/dp_\text{T}^2$ denotes the yield of charged particles normalized by $2\pi p_\text{T}$, the number of events, and pseudorapidity interval. 
Nevertheless, some experiments report the spectrum without the $p_\text{T}$ normalization, usually denoted by $dN/dp_\text{T}$. In such cases, Eq.~\eqref{eq:U}, and the following presented in this manuscript, must be multiplied by $p_\text{T}$ to fit the data correctly, keeping all the results discussed in this paper valid.

\section{Elucidating the softened hadron production}
\label{sec:elucidating}

\begin{figure}
    \centering
    \includegraphics{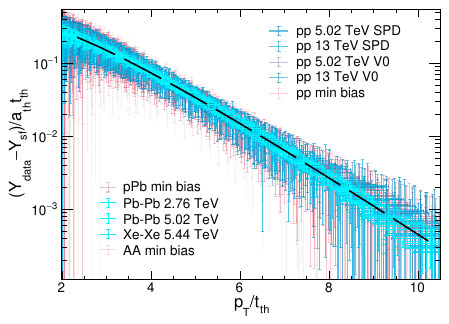}
    \caption{Difference of experimental yield data ($Y_\text{data}$) and the interpolated yield $(Y_\text{sf})$ normalized by $a_{th}t_{th}$ as a function of $p_\text{T}/t_{th}$ for all the experimental analyzed data at intermediate $p_{\text{T}}$ region. Error bars are the uncertainty propagation of the experimental data together with the fitting parameter errors.
    The dashed line corresponds to the function $xe^{-x}$.}
    \label{fig:sustento}
\end{figure}

We assess the contribution of softened hadrons by analyzing the experimental data of pp, AA, and pPb collisions at different center of mass energies, centralities, and multiplicity classifications. 
To this end, we assume that the contributions of soft and hard scales mainly come from the string fragmentation picture. As a first step, we search for the best fit of the $U$ function \eqref{eq:U} to data by excluding the intermediate $p_\text{T}$ region. Therefore, we interpolate the yield produced by the string fragmentation ($Y_\text{sf}$) in the intermediate region by evaluating the bin centers onto the $p_\text{T} U$ function.
We found that the difference of the experimental yield data ($Y_{\text{data}}$) and $Y_\text{sf}$ behaves as $Y_{\text{data}} - Y_{\text{sf}} =  a_{th} p_\text{T} e^{-p_\text{T}/t_{th}}$ in the range of 1 to 6 GeV for all data sets analyzed in this manuscript. In Fig.~\ref{fig:sustento}, we plot $(Y_\text{data}- Y_\text{sf})$  normalized by $a_{th} t_{th}$ as a function of the transverse momentum scaled by $t_{th}$. 
Note that we use $Y$ to denote the yield $dN/dp_\text{T}$ to improve notation.
Remarkably, all the analyzed data at intermediate $p_\text{T}$ values collapse following a universal behavior. 
We must emphasize that this fact does not imply that the production of softened hadrons comes from thermal processes, but these particles redistribute their transverse momentum in this form.

\begin{figure*}
    \centering
    \includegraphics{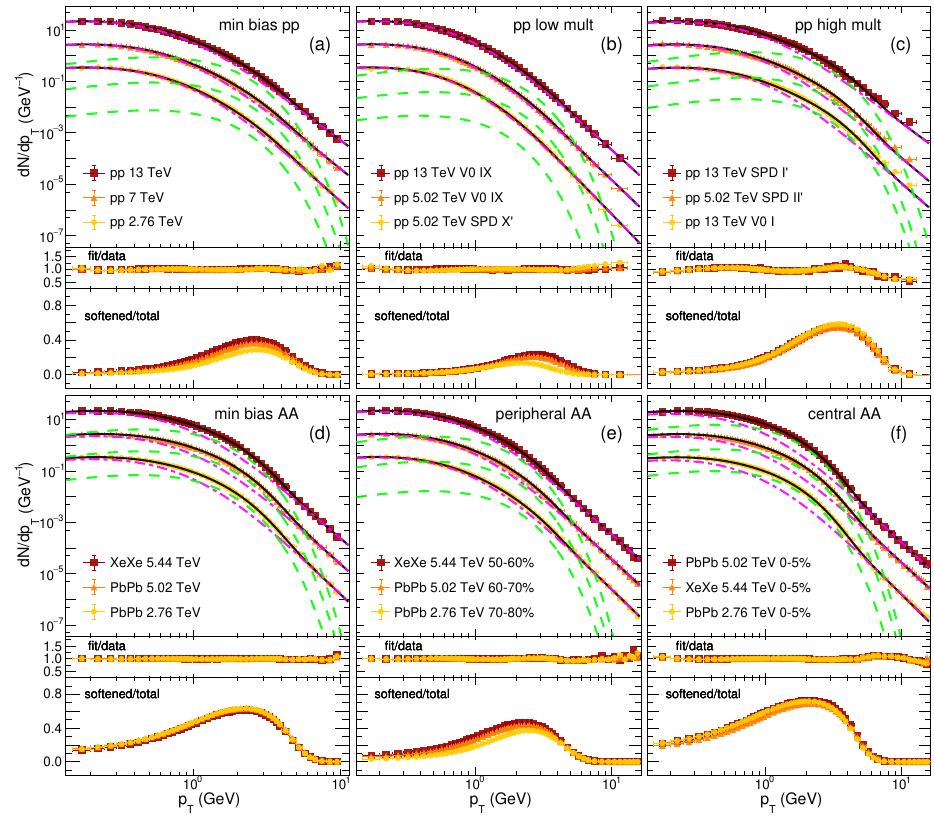}
    \caption{Sample of the $dN/dp_\text{T}$ data sets analyzed for pp collision: (a) minimum bias, (b) low multiplicities, and (c) high multiplicities. For PbPb and XeXe collisions: (d) minimum bias, (e) peripheral, and (d) central.
    The fit to data and softened hadrons' contribution to total charged particle production ratios are plotted for each case.
    The $dN/dp_\text{T}$ data sets are plotted with marks. Solid lines correspond to the total fit~\eqref{eq:fullTMD}. Dash-dotted and dashed lines are the Tricomi's function and $p_\text{T}$-exponential contributions to the total charged particle production, respectively. We scaled all data sets to improve visualization.}
    \label{fig:Fits}
\end{figure*}

\section{Description of the complete $p_\text{T}$-spectrum}
\label{sec4:complete}

Supported by the previous result on the soft hadron production discussed in Sec.~\ref{sec:elucidating}, we propose to add a $p_\text{T}$-exponential term in Eq.~\eqref{eq:U} to obtain a complete description of the $p_\text{T}$-spectrum, i.e., 
\begin{equation}
 \frac{dN}{d p_\text{T}^2} = A_U U\left(a,1/2,z_0 p_\text{T}^2\right) + A_\text{th} e^{-p_\text{T}/T_\text{th}, } 
 \label{eq:fullTMD}
\end{equation}
where $T_{th}$ corresponds to the $p_\text{T}$ scale related to the softened hadron production.
We fit Eq.~\eqref{eq:fullTMD} to the $p_\text{T}$-spectrum data at midrapidity reported in Refs.~\cite{ALICE:2022xip,ALICE:2019dfi,ALICE:2018vuu,ALICE:2018hza} by the ALICE Collaboration for pp, pPb, XeXe, and PbPb collisions at LHC energies considering V0, SPD, and centrality classifications. In Eq.~\eqref{eq:fullTMD}, the free parameters $A_U$ and $A_{\text{th}}$ can be understood as the weights for the contribution of the particles produced by the string fragmentation and softened hadrons to the total yield, respectively. 

\begin{figure*}
    \centering    \includegraphics{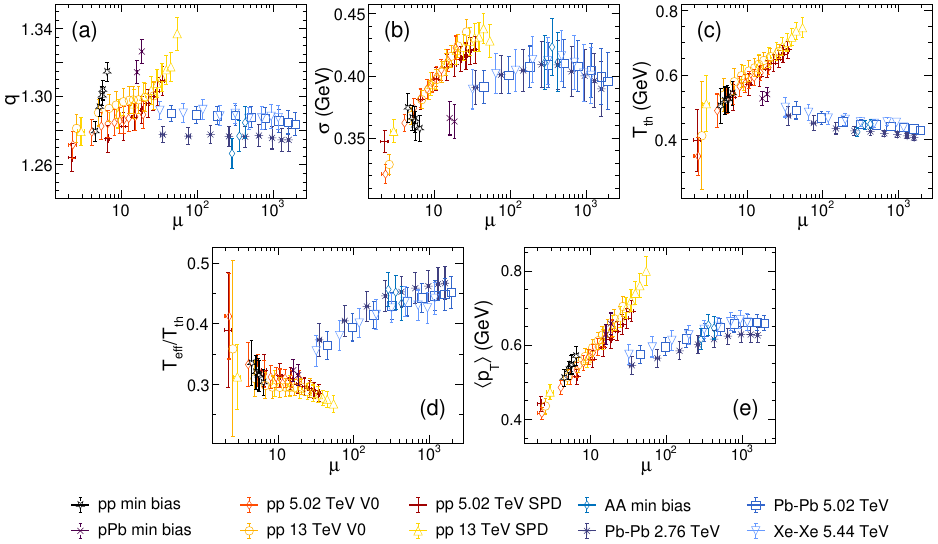}
    \caption{Multiplicity ($\mu$) dependence of the fitting parameters (a) $q$, (b) $\sigma$, (c) $T_\text{th}$, (d) the $T_\text{eff}/T_\text{th}$ ratio, and (e) $\langle p_\text{T} \rangle$ for all analyzed data.}
    \label{fig:fitpar}
\end{figure*}

In Fig.~\ref{fig:Fits}, we show samples of fits of Eq.~\eqref{eq:fullTMD} to $dN/dp_\text{T}$ data sets of pp collisions [(a) minimum bias, (b) low and (c) high multiplicity events], and AA collisions [(d) minimum bias, (e) peripheral and (d) central events].
In all cases, we show the contribution fraction to the total yield of the functions $p_\text{T} U$ (pink dash-dotted line) and $p_\text{T} e^{-p_\text{T}/T_\text{th}}$ (green dashed line). 
In particular, the highest values of the ratio of softened hadrons contribution to total yield indicate the $p_\text{T}$ bins wherein the softened hadrons become relevant, which occurs at the intermediate $p_\text{T}$ region. 
Additionally, we found that $\chi^2/\text{ndf} < 1$ for all the analyzed data.

\begin{figure*}
    \centering   
      \includegraphics{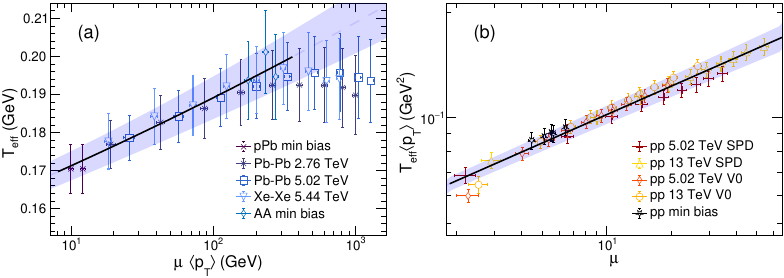}    
   \caption{(a) Effective temperature as a function of $\mu \langle p_\text{T} \rangle$ for pPb and AA collisions systems, where $\mu$ denotes the multiplicity. The dashed line corresponds to the extrapolation of the data trend for $\mu \langle p_\text{T} \rangle > 300$ GeV. (b) $T_\text{eff} \langle p_\text{T} \rangle$ as a function of the multiplicity ($\mu$) for pp collisions. Shaded regions correspond to the 2-$\sigma$ uncertainty propagation. }
    \label{fig:Tes}
\end{figure*}

On the other hand, Figures~\ref{fig:fitpar} (a)-(c) show the value of the fitting parameters $q$, $\sigma$, and $T_\text{th}$ as a function of the multiplicity for the data sets analyzed, respectively. These parameters allow us to characterize the properties of the systems through the $p_\text{T}$-spectrum. For instance, in the limit of low $p_\text{T}$, Eq.~\eqref{eq:fullTMD} resembles the thermal distribution with effective temperature:
\begin{equation}
   T^{-1}_\text{eff} =  \frac{1}{ A_U^* + A_{th} }   \left(    \frac{A_U^*}{T_U}    +   \frac{A_{th}}{T_{th}}        \right), 
\label{eq:Teff}
\end{equation}
where $A_U^* = \sqrt{\pi}A_U/\Gamma( a+ 1/2) $. 
The effective temperature $T_{\text{eff}}$ can be understood as the soft scale linked to the complete $p_\text{T}$-spectrum, considering the contribution of charged particle production coming from string fragmentation and collective phenomena. This is computed over the ensemble of collision events occurring under the same biases \cite{Garcia:2023eqg}.
Notice that the definition of $T_{\text{eff}}$ in Eq.~\eqref{eq:Teff} is valid for all systems and it is computed by using the fitting parameters extracted from analyzing the $p_\text{T}$-spectrum. 
We expect that $T_{\text{eff}}$ will exhibit some system-size dependence because the $p_\text{T}$-spectrum behaves and evolves differently with multiplicity for pp and AA collisions. 
However, in all cases, we found that $T_{\text{eff}}<T_\text{th}$ (see Fig.~\ref{fig:fitpar} (d)).
It means that the softened hadrons take $p_\text{T}$ values beyond the soft scale and enhance the spectrum at the intermediate $p_\text{T}$ region, which happens at larger $p_\text{T}$ values for pp than heavy ion collisions. This is a consequence of the larger values of $T_\text{th}$ for pp than AA (see Fig.~\ref{fig:fitpar} (c)).
The shift of the $p_\text{T}$ scale where the softened hadrons' contribution takes its maximum values in pp and AA collisions is consistent with the system's size effects, as shown in the ratio of the softened hadrons' contribution to the total charged particle production plotted in Fig.~\ref{fig:Fits}.

Another interesting asymptotic limit occurs at high-$p_\text{T}$ values. In this limit, Eq.~\eqref{eq:fullTMD} behaves as the power law  $(p_\text{T}^2)^{\frac{1}{2}-\frac{1}{q-1}}$ because the contribution of the $p_\text{T}$-exponential term vanishes in this limit. Notice that the relevance of $q$ lies in describing the shape of the $p_\text{T}$-spectrum tail. Lower $q$ values mean a decrement in the probability of producing high $p_\text{T}$ hadrons. 
In fact, for AA collision, we found a saturation on $q$, and it slightly decreases at the largest multiplicity classes (see Fig.~\ref{fig:fitpar} (a)), meaning the suppression of the high-$p_\text{T}$ particle production in the most central collisions. This effect is aligned with the jet-quenching phenomena reported in heavy ion collisions. 
In these large systems, this behavior is related to the energy loss, which mainly affects partons originated from deep scattering processes,  resulting in a suppression of high-$p_\text{T}$ hadrons \cite{Bjorken:1982tu,PHENIX:2001hpc,STAR:2002ggv}. 
Conversely, in pp collisions, the suppression of high-momentum hadrons is expected to be lesser due to the system size. 
Notice that our findings of the contribution of the softened hadrons agree with those observations. 

In addition, the computation of the average $p_\text{T}$ from the $p_\text{T}$-spectrum must take into account the extra $p_\text{T}$ from the phase space factor, i.e., 
\begin{equation}
  \langle p_\text{T} \rangle = \frac{\int_0^\infty p_\text{T}^2 \frac{dN}{dp_\text{T}^2} dp_\text{T}}{\int_0^\infty p_\text{T} \frac{dN}{dp_\text{T}^2} dp_\text{T} }. 
\end{equation}
Then, for the $p_\text{T}$-distribution given by \eqref{eq:fullTMD}, we found
\begin{equation}
\langle p_\text{T}\rangle =
\frac{ A_U I_1 \langle p_\text{T}\rangle_U +  2A_\text{th}T_\text{th}^3 }{ 
 A_U I_1 +  A_\text{th} T_\text{th}^2},
\label{eq:meanpT}
\end{equation}
where $$I_1=\int_0^\infty p_\text{T} U dp_\text{T},$$ $$A_U I_1=A_U^* \sigma^2 /\pi(5-3q),$$ and $$\langle p_\text{T}\rangle_U= \sigma \sqrt{2} \Gamma(a-3/2)/\sqrt{ \pi (q-1)} \Gamma(a-1),$$ which converges if $1<q< 3/2$ \cite{Herrera:2024zjy}. 
Notice that we have computed the average $p_\text{T}$ by integrating over $[0,\infty)$. Figure~\ref{fig:fitpar} (e) contains our results of the $\langle p_\text{T} \rangle$ calculations.
However, we reproduce the values of $\langle p_\text{T} \rangle$ by numerically integrating over the $ p_\text{T}$ range set up by the ALICE Collaboration reported in Ref. \cite{ALICE:2022xip}.

We found that the effective temperature shows different behaviors in small and large systems. 
For instance, $T_{\text{eff}}$ scales as $(\mu \langle p_\text{T} \rangle)^{\beta_1}$ in the cases of AA and pPb collisions, where $\beta_1$=0.043(2) and $ \mu$ denotes the average multiplicity of charged particles per pseudorapidity interval.
The factor $\mu \langle p_\text{T} \rangle$ is relevant because it is related to the Bjorken energy \cite{Bjorken:1982qr}, which can now be rewritten only in terms of the effective temperature.
Notice that we are referring to Bjorken energy instead of the Bjorken energy density. The latter needs to be divided by the transverse area and proper time.
Conversely, as the multiplicity increases, $T_\text{eff}$ saturates for $\mu \langle p_\text{T} \rangle>300 $ GeV (see Fig~\ref{fig:Tes} (a)), 
also observed in other analyses of experimental data \cite{STAR:2008med,Feal:2018ptp}. 
On the other hand, for pp collisions, we found that the effective temperature relates to multiplicity and  $\langle p_\text{T} \rangle$ through $T_\text{eff} \propto \mu^{\beta_2} /\langle p_\text{T} \rangle$, with $\beta_2$=0.270(9), as shown in Fig.~\ref{fig:Tes} (b). 
We must emphasize that our findings reveal two distinct behaviors of the effective temperature depending on the kind of the analyzed system. These behaviors are solid evidence of the differences between the systems formed in small and large collisions and their posterior evolution.

\section{Softened hadrons' contribution to the $p_\text{T}$-spectrum, $N_{\text{ch}}$ and $\langle p_\text{T} \rangle$}
\label{sec:contribution}

\begin{figure*}
    \centering
      \includegraphics{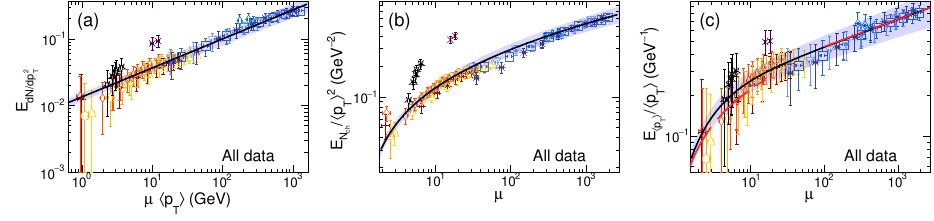}    
   \caption{(a) Softened hadrons contribution to the $p_\text{T}$-spectrum as a function of $\mu \langle p_\text{T} \rangle$ for all the analyzed cases. (b) The fraction of the total yield composed by the softened hadrons as a function of the multiplicity. 
   (c) Dependence of $\text{E}_{\langle p_\text{T}\rangle}/\langle p_\text{T}\rangle$ on the multiplicity for all the analyzed cases. The dashed lines show the asymptotic behaviors of the scaling law at low and high multiplicities. 
   In all panels, solid lines correspond to the scaling laws. Shaded regions correspond to the 2-$\sigma$ uncertainty propagation. }
    \label{fig:scalingL}
\end{figure*}

To quantify the contribution of the softened hadrons to the physical observables computed from the $p_\text{T}$-spectrum, we define
\begin{equation}
\text{E}_n
           =\frac{A_\text{th}\int_0^\infty p_\text{T}^n e^{-p_\text{T}/T_\text{th}} dp_\text{T}}{\int_0^\infty p_\text{T}^n  \frac{dN}{dp_\text{T}^2} dp_\text{T}}
           = \frac{ A_\text{th} n! T^{n+1} }{A_U I_n + A_\text{th} n! T^{n+1} },
           \label{eq:Excn}
\end{equation}
where $I_n=\int_0^\infty p_\text{T}^nUdp_\text{T}$ is well defined if $q<(4+n)/(2+n)$ \cite{Herrera:2024zjy}.

In particular, the cases $n=0,1,2$ correspond to the weighted contribution of softened hadrons to the $p_\text{T}$-spectrum, number of charged particles produced ($N_{\text{ch}}$), and $\langle p_\text{T} \rangle$, respectively. Note that for $p_\text{T}$-spectra with no softened hadron production $\text{E}_n =0$, indicating all the particle production comes from the fragmentation of color strings. 
However, as the mechanisms beyond the string fragmentation appear, the hadrons' $p_\text{T}$ is redistributed in a particular way that increases the $\text{E}_n$ values.

The contribution of the softened hadrons to the $p_\text{T}$-spectrum is quantified by Eq.~\eqref{eq:Excn} for $n=0$, denoted by $ \text{E}_{dN/dp_\text{T}^2}$.
We found that $\text{E}_{dN/dp_\text{T}^2}\propto (\mu \langle  p_\text{T} \rangle)^{\beta_3}$, with $\beta_3$=0.43(1) (see Fig.~\ref{fig:scalingL} (a)). Interestingly, this relation stands for the $p_\text{T}$-spectrum data from pp to AA collisions, indicating that the production of hadrons with intermediate $p_\text{T}$ increases with the midrapidity Bjorken energy \cite{Bjorken:1982qr,STAR:2008med}.

The fraction of the total yield composed by the softened hadrons is computed using Eq.~\eqref{eq:Excn} for $n=1$,  denoted by $ \text{E}_{N_{\text{ch}}}$.
Figure~\ref{fig:scalingL} (b) shows the behavior of $\text{E}_{N_{\text{ch}}}/\langle p_\text{T} \rangle^2$ as a function of multiplicity. 
We found that this ratio scales for all the analyzed data as $\mu^{\beta_{4}}$, with $\beta_{4} = 0.08(1)$.
Notice that $\text{E}_{N_{\text{ch}}}$ vanishes for the lowest multiplicity events. 
In fact, in the lowest multiplicity classes of pp collisions, our results indicate that the softened hadron production is, on average, one particle per 50 events, which assures that the production of charged hadrons is due to the fragmentation of color strings in this limit.
This means the probability of producing this kind of particle is very low but grows with multiplicity. 
For comparison, the softened hadron production for the lowest multiplicity events in pp collisions at 13 TeV is two orders of magnitude lower than for the cases of the highest multiplicity classes.
Interestingly, the cases of the highest multiplicity pp collisions have comparable values of the fraction of the total yield of peripheral AA collisions.
Moreover, for the most central AA collisions, $\text{E}_{N_\text{ch}}$ takes its maximum value, and the total yield is composed of around 30\% of softened hadrons.

The softened hadrons' contribution to $\langle p_\text{T} \rangle$ is given by Eq.~\eqref{eq:Excn} for $n=2$, denoted by $\text{E}_{\langle p_\text{T}\rangle}$.
In Fig.~\ref{fig:scalingL} (c), we plot the ratio $\text{E}_{\langle p_\text{T}\rangle}/\langle p_\text{T}\rangle$ as a function of the multiplicity, which collapses following a monotone increasing trend for all the analyzed cases. Interestingly, the data of the ratio $\text{E}_{\langle p_\text{T}\rangle}/\langle p_\text{T}\rangle$ can be well described by the scaling relation $\text{E}_{\langle p_\text{T}\rangle}/\langle p_\text{T}\rangle \propto \mu^{\beta_5}+e^{-2/\mu}-1$,  with $\beta_5$=0.20(1). 
The latter has two asymptotic behaviors. In the case of $\mu \to 1$, we found $\text{E}_{\langle p_\text{T}\rangle}/\langle p_\text{T}\rangle \propto \ln \mu$. On the other hand, for events with high multiplicity, $\text{E}_{\langle p_\text{T}\rangle}/\langle p_\text{T}\rangle \propto \mu^{\beta_5}$.
Even though we have found a universal scaling law that includes the data of pp, pPb, and AA collisions, the two distinct asymptotic behaviors may indicate size effects for small and large systems in producing the softened hadrons.
In particular, the large production of softened hadrons for the highest multiplicity events in AA collisions leads to a saturation and a possible reduction of the average $p_\text{T}$ (see Fig.~\ref{fig:fitpar} (e)) \cite{ALICE:2022xip}, which is consistent with large effects from collective phenomena that degrade the $p_\text{T}$ of the hard hadrons, which later produce particles at a lower $p_\text{T}$.

To close this section, let us comment on the results associated with the minimum bias cases, whose results set them apart as outliers. This may happen since the minimum bias data contains the total information from all the multiplicity classes. Although high multiplicity events drive the shape of the $p_\text{T}$ spectrum tail and enhance the production of softened hadrons, their contributions to the freeze-out parameters are lower than those provided by the low multiplicity events.

\section{Conclusions}
\label{sec:conclu}

In summary, we described the experimental $p_\text{T}$-spectrum data through Tricomi's function by assuming that the fragmentation of color strings is a baseline for the charged particle production.
However, this approach deviates from data at intermediate $p_\text{T}$ region in the cases of the most energetic collisions, highest multiplicities, and central AA collisions. We assumed that such deviation is originated by the production of softened hadrons coming from processes beyond string fragmentation, such as color reconnection, parton recombination, jet quenching, energy loss, and flow, among other collective phenomena.
Through a data-driven study, we found that such deviations follow a universal behavior that collapses into a $p_\text{T}$-exponential decay in a range of 1 GeV to 6 GeV, which stands for the experimental data of pp, pPb, PbPb, and XeXe collisions at LHC energies.
Supported by this insight, we describe the entire $p_\text{T}$-spectrum of charged particle production by incorporating a $p_\text{T}$-exponential term to the Tricomi's function (see Eq.~\eqref{eq:fullTMD}). Then, we analyzed the $p_\text{T}$-spectrum data reported by the ALICE Collaboration concerning the production of charged particles in ultrarelativistic collisions from pp to AA, considering different biases, namely minimum bias, multiplicity, and centrality classifications \cite{ALICE:2022xip,ALICE:2019dfi,ALICE:2018vuu,ALICE:2018hza}. 
Remarkably, Eq.~\eqref{eq:fullTMD} accurately describes the $p_\text{T}$-spectrum data for all the analyzed cases. We recall that the Tricomi's function is deduced from a nonextensive description of the string tension fluctuations, which is inherited by Eq. \eqref{eq:fullTMD}. Thus, it preserves the power law tail behavior in the limit of high $p_\text{T}$ values.
Note that the additional $p_\text{T}$-exponential enhances the soft part, reaching its maximum contribution at intermediate $p_\text{T}$ but vanishes at high $p_\text{T}$. In fact, the inverse effective temperature is computed as a weighted sum of $T_U^{-1}$ and $T_\text{th}^{-1}$.
We must emphasize that the $p_\text{T}$-exponential term in Eq. \eqref{eq:fullTMD} does not mean that thermal processes produce the softened hadrons. However, the aforementioned mechanisms redistribute the transverse momentum in this particular way.

We found an enhancement of the $p_\text{T}$-spectrum for intermediate $p_\text{T}$ values, which is more noticeable as the multiplicity increases (see Fig.~\ref{fig:Fits}). It is consistent with similar signals for QGP formation reported by other analyses, such as the two particle angular correlations and collective flow effects \cite{CMS:2010ifv,CMS:2016fnw}.
In the cases of heavy ion collision, we observed that the maximum contribution of the softened hadrons to the $p_\text{T}$-spectrum occurs for $p_\text{T}$ values below than the case of pp collisions. This difference is expected because of the system's size effects: the larger the systems are, the more relevant suppression of high $p_\text{T}$ hadrons.

Another evidence of the system's size effects is observed in the effective temperature, which behaves differently for small and large systems.
In pp collisions, the combination $T_\text{eff}\langle p_\text{T} \rangle $ data collapses onto a power law trend with variable $\mu$. 
In the cases of pPb and AA collisions, $T_\text{eff}$ grows as a power law in the variable $\mu\langle p_\text{T} \rangle $, which occurs for $\mu < 600$ or $\mu\langle p_\text{T} \rangle < 300$ GeV, saturating for larger $\mu$ values. Note that $\mu\langle p_\text{T} \rangle $ is proportional to the Bjorken energy, which now can be rewritten in terms only of the effective temperature.

We also found that the contribution of the softened hadrons to the $p_\text{T}$-spectrum and $\langle p_\text{T} \rangle$ can collapse in global trends. 
It is worth mentioning that our results indicate a close relation between $\text{E}_{dN/dp_\text{T}^2}$ and the Bjorken energy. 
Moreover, the ratio $\text{E}_{\langle p_\text{T} \rangle} / \langle p_\text{T} \rangle$ as well as $\text{E}_{N_\text{ch}} / \langle p_\text{T} \rangle^2$ only depend on the multiplicity. 
Our results on $\text{E}_{dN/dp_\text{T}^2}$ and $\text{E}_{\langle p_\text{T} \rangle}$ are consistent with other analyses claiming collective effects in both pp and AA collisions \cite{ALICE:2015wav,Kalaydzhyan:2015xba}.

The model presented in this manuscript is helpful for quantifying the fraction of softened hadrons with intermediate $p_\text{T}$ values and their contribution to the $p_\text{T}$-spectrum, effective temperature, and $\langle p_\text{T} \rangle$. 
The main results presented in this manuscript are the scaling laws for the contribution of the softened hadrons to the $p_\text{T}$-spectrum and average $p_\text{T}$. From this, we can conclude that the physical mechanisms producing the charged softened hadrons have similar origins for all colliding systems.
However, the approach presented here cannot distinguish the mechanisms for producing the softened hadrons. 
Nevertheless, by using our methodology, it is possible to estimate the composition of the softened hadrons by computing $\text{E}_{N_\text{ch}}$ for the yield of identified particles.
Additionally, our results can be helpful in improving color string based models and computational tools used to simulate high energy collisions.

\section*{Acknowledgments}
This work was funded by Consejo Nacional de Humanidades, Ciencias y Tecnologías (CONAHCYT-México) under the project CF-2019/2042,
graduated fellowship grant number 1140160, and postdoctoral fellowship grant numbers 645654 and 289198, and  DGAPA-PAPIIT BG100322 project.

\bibliography{ref.bib}

\end{document}